\documentclass[aps,prx,reprint,superscriptaddress]{revtex4-1}

\usepackage[version=3]{mhchem}
\usepackage{bm}
\usepackage{amsmath}
\usepackage{ulem}
\usepackage{amssymb}
\usepackage{amsfonts}
\usepackage{euscript}
\usepackage{color}
\usepackage{epsfig}
\usepackage{hyperref}
\usepackage[section]{placeins}
\usepackage{gensymb}
\usepackage{array}

\makeatletter
\newcommand{\thickhline}{
    \noalign {\ifnum 0=`}\fi \hrule height 1pt
    \futurelet \reserved@a \@xhline
}
\newcolumntype{"}{@{\hskip\tabcolsep\vrule width 1pt\hskip\tabcolsep}}
\makeatother

\begin{document}

\title
  {Circular dichroism induced by Fano resonances\\ in planar chiral oligomers}

\author{Ben~Hopkins}
\affiliation{Nonlinear Physics Centre, Australian National University, Canberra, ACT 2601, Australia}
\email{ben.hopkins@anu.edu.au}

\author{Alexander~N.~Poddubny}
\affiliation{Ioffe Physical-Technical Institute of the Russian Academy of Sciences, St. Petersburg 194021, Russia}
\affiliation{National Research University for Information Technology, Mechanics and Optics, St. Petersburg 197101, Russia}

\author{Andrey~E.~Miroshnichenko}
\affiliation{Nonlinear Physics Centre, Australian National University, Canberra, ACT 2601, Australia}

\author{Yuri~S.~Kivshar}
\affiliation{Nonlinear Physics Centre, Australian National University, Canberra, ACT 2601, Australia}
\affiliation{National Research University for Information Technology, Mechanics and Optics, St. Petersburg 197101, Russia}

\graphicspath{{./Figures/}}

%%%
%\setpages[]{}
%\setvolume[0]{0}
%\setyear{2011}%
%\setdoi{201100000}%
%\oddheadlogotext{}{Guide}
%%% 

\begin{abstract}
We present a general theory of circular dichroism in planar chiral nanostructures with rotational symmetry.
It is demonstrated, analytically, that the handedness of the incident field's polarization can control whether a nanostructure induces either absorption or scattering losses, even when the total optical loss (extinction) is polarization-independent. 
We show that this effect is a consequence of modal interference so that strong circular dichroism  in absorption and scattering can be engineered by combining Fano resonances with planar chiral nanoparticle clusters.
\end{abstract}

\title{Circular dichroism induced by Fano resonances in planar chiral oligomers}

\maketitle

\section{Introduction}

The difference in absorption of left- and right-handed circularly-polarized (LCP and RCP) light by chiral structures has long been utilized for applications in molecular chemistry~\cite{Fasman1996,Barron2004}, pharmaceuticals~\cite{Hutt1996, Nguyen2006}, and optics~\cite{Kitzerow2001, Pendry2004}. 
More recently, advances in nanotechnology have resulted in new types of chiral nanoantennas, colloidal nanoparticles, and metamaterials; nanostructured systems which provide unprecedented freedom to produce chiral responses that substantially exceed those of conventional materials.{~\cite{Wang2009, Soukoulis2011, Hildreth2012, Maksimov2014, Cui2014, Zhu2014,Lobanov2015}}
{It is, however, {\it planar} nanostructures that} have the least restrictions in design, particularly when considering fabrication, and, therefore, they are more suitable to capitalize on design freedom and produce the most effectual chiral responses.
{Yet, such two-dimensional nanostructures cannot be truly chiral because they have an inherent plane of mirror symmetry. 
In combination with reciprocity, this is known to prevent any difference in the total optical losses under LCP or RCP incident fields -- a difference known as {\it circular dichroism}.}
 Here, however, we consider separate contributions from the radiative and dissipative components of a system's total loss.   
As we show, these components of the total loss, the scattering and absorption cross-sections, are not constrained by reciprocity and can subsequently exhibit circular dichroism in planar chiral geometries and metasurfaces (see Fig.~\ref{fig:schematic}). 
Here we focus on circular dichroism in the absorption cross-section. 
This definition is warranted from a practical perspective as, for instance, the absorption cross-section will encompass nonlinearity~\cite{Valev2009}, photocurrent generation~\cite{Fan2014}, heating~\cite{Carlson2012}, fluorescence~\cite{Swasey2014} and photocatalysis~\cite{Mukherjee2013}; quantities that are both measurable and foreseeably useful.

{\begin{figure*} [ht!]
\centerline{\includegraphics[width=0.7\textwidth]{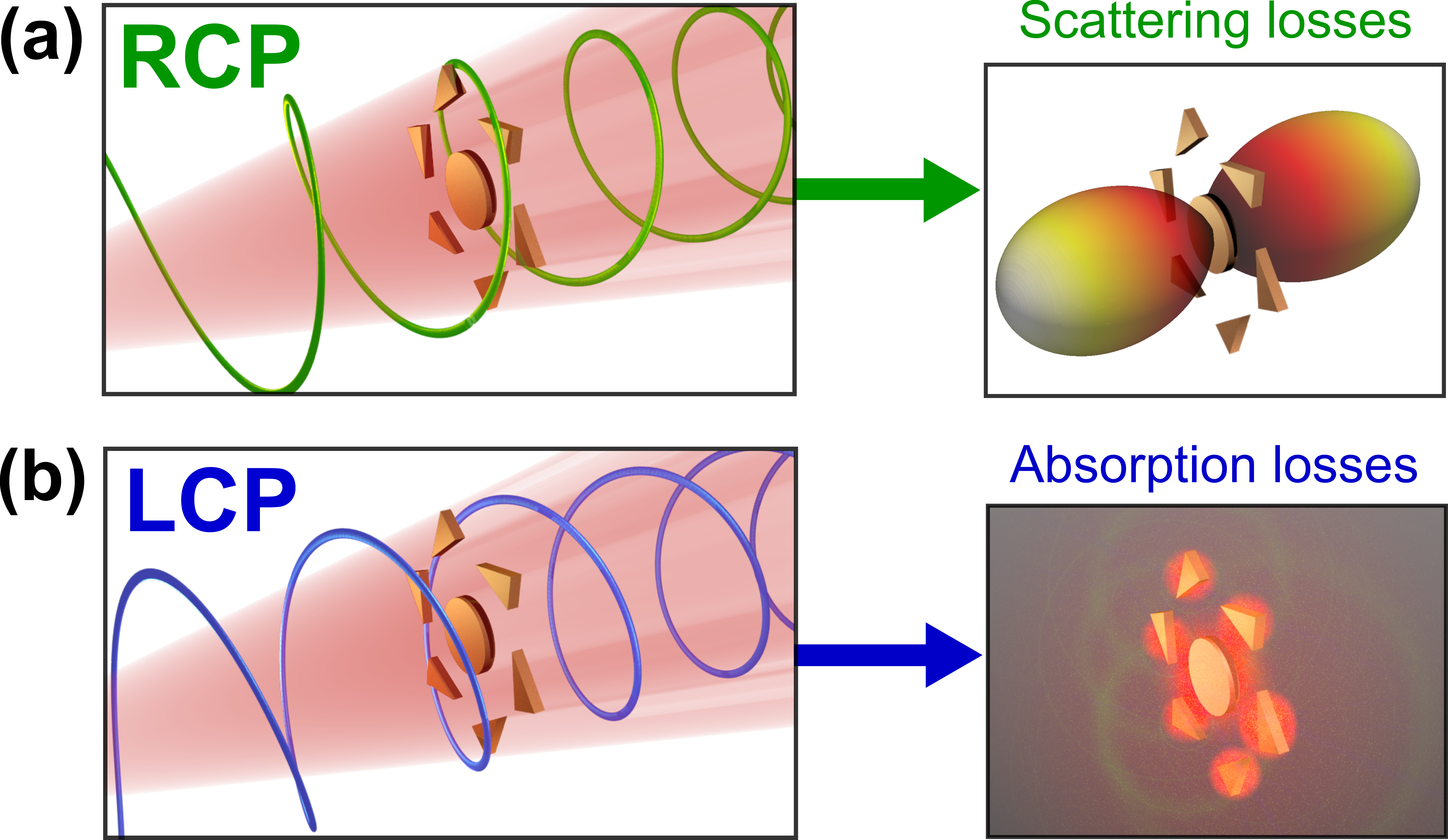}}
\caption{A schematic of the circular dichroism in absorption observed for planar structures excited by a normally incident plane wave.  The total energy loss (extinction) by a planar chiral structure is conserved due to reciprocity. But, the particular loss mechanism can be swapped between (a) scattering and (b) absorption with the handedness of the incident field polarization. }
\label{fig:schematic}
\end{figure*}  }

The effect of chirality on scattering and absorption is generally understood by how it impacts either the far-field scattering matrices or the effective medium parameters of a given structure~\cite{Bai2007, Zhao2010}.
However, these two approaches have been tailored (historically) towards characterizing the chiral properties of conventional materials and do not provide information on what is physically occurring in the system.
This means that it can be highly nontrivial to deduce how a general change in geometry will effect the far-field scattering matrices or effective parameters. In an attempt to address this, some recent works have considered the properties of the near-field of certain chiral structures to guide design empirically~\cite{Schaferling2012} or attempted to describe the response of chiral arrangements of nanorod dimers analytically~\cite{Chigrin2011, Davis2012}.
Note here, the relation between the chiral response and the geometry is ultimately rooted in the current distribution that is induced by the given incident field.  Specifically, these currents will encapsulate the complete optical response of any system in both the near- and far-field regions, and should thereby provide a deeper understanding of the observed optical responses. The approach we present here will be to simplify the analysis of structures using geometric symmetry rather than analytical approximations. In this regard, there have been significant works on planar chiral metasurfaces and metamaterials whose constituent meta-atoms have a discrete rotational symmetry~\cite{Plum2009,Decker2010}.
A number of key advantages of utilizing rotational symmetry in applications that require polarization-selective operation have further been proposed.  In particular: invariance of scattering and absorption loss to all linear polarizations\footnote{This is necessary to remove the cross-section's dependence on the phase between LCP and RCP light.}~\cite{HopkinsLiu2013}, zero transmitted circular cross polarization~\cite{Fernandez-Corbaton2013} and others~\cite{Lubin2012, Rahmani2013, Chen2014, Konishi2014}.
Therefore, while imposing symmetry conditions necessarily restricts the range of chiral nanostructures that can be considered for a given application, the remaining subset of nanostructures (which are both chiral and rotationally-symmetric at the same time) are desirable and have been just as effective at exhibiting chiral optical responses. 
We have subsequently focused here on symmetries that are rotationally symmetric and lack any mirror plane parallel to the rotational symmetry axis. 
We do not consider geometries that have a mirror plane parallel to their rotational symmetry axis, such as in the $C_{nv}$ and $D_{nh}$ point {groups. 
This is because such mirror symmetry operations are able to transform an LCP plane wave into an RCP plane wave (and {\it vice versa}), thereby leaving the geometry unchanged when flipping the handedness of the excitation.  
This would subsequently make the optical responses to LCP and RCP plane waves symmetrically-equivalent.
We therefore consider} the following symmetry groups: $C_n$, $D_n$, $S_n$ and $C_{nh}$ (see Fig.~\ref{fig:symmetries}). 
While the $C_n$ and $D_n$ point groups are chiral, it is important to acknowledge that $S_n$ and $C_{nh}$ point groups are, in fact, {\it achiral}; a result of their symmetry under improper rotation and planar mirror reflection operations, respectively.  
We focus on the $C_{nh}$ point group in the main text, corresponding to geometries that are chiral in only two dimensions, a property which is often referred to as {\it planar chiral}.
We also consider only the incident waves propagating along the symmetry axis.
{\begin{figure} [h!]
\centerline{\includegraphics[width=0.95\columnwidth]{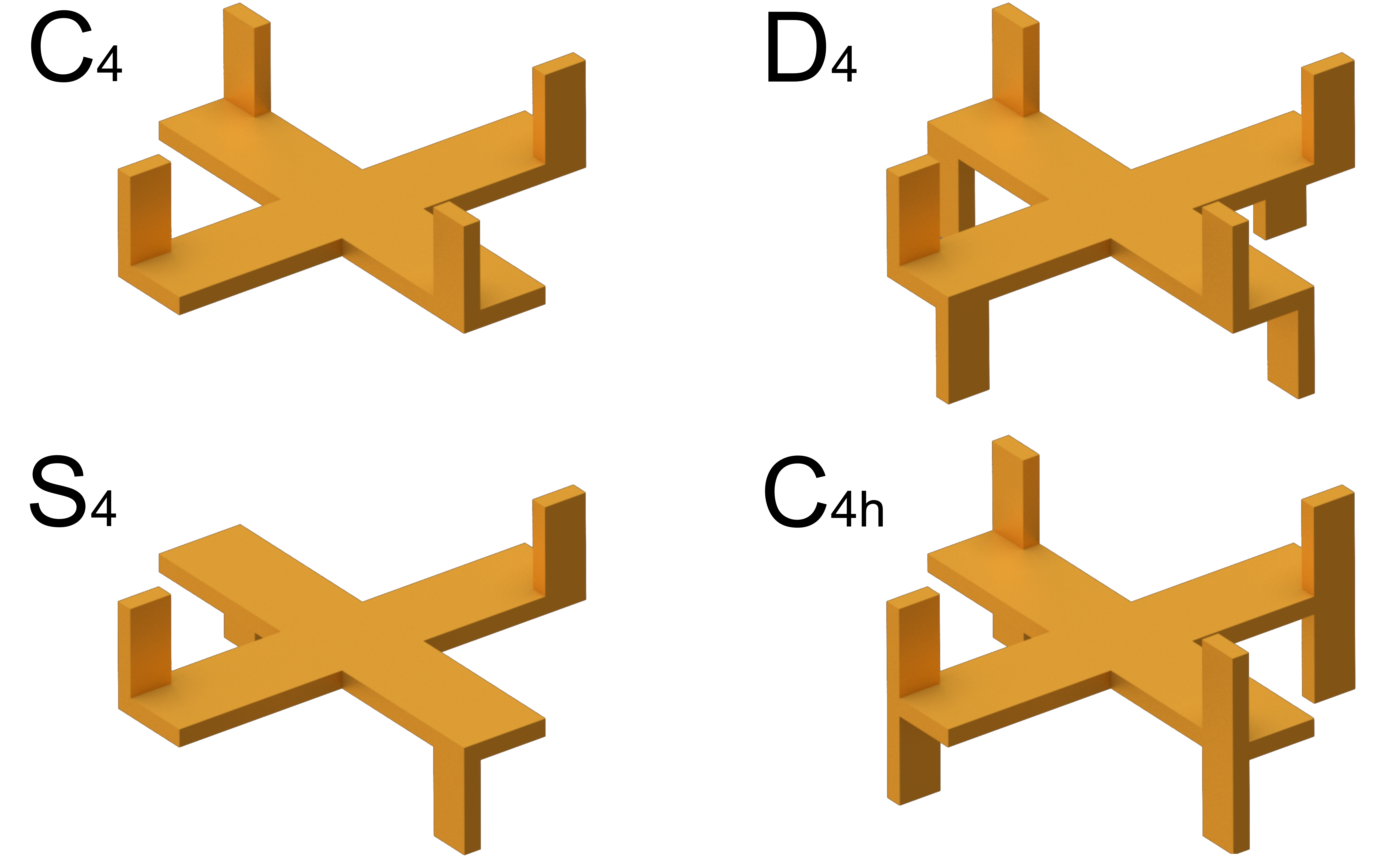}}
\caption{Examples of geometries that have discrete rotational symmetry, specifically (clockwise the top left): cyclic group $C_n$, dihedral group $D_n$,  cyclic group with a horizontal mirror symmetry $C_{nh}$ and symmetric group $S_n$.  These examples are for $n=4$, however our approach is applicable to any geometry where $n\geq3$. }
\label{fig:symmetries}
\end{figure}  }
In this paper, we demonstrate how symmetry considerations can allow us to analytically extricate meaningful information from the currents induced in { $C_{nh}$} structures without calculating the explicit current distributions.
From this foundation, we are able to link the origin of non-reciprocal circular dichroism in the absorption cross-section to modal interference and Fano resonances.

\section{Reciprocity and chirality}

To provide broadly-applicable arguments about the induced currents in any geometry, we will work from frequency-domain Maxwell's equations, in the absence of magnetization and magnetic sources for simplicity. Let our scattering structure be some localized distribution of a material having a permittivity not equal to the background permittivity ($\epsilon_0$), in which a current distribution ($\boldsymbol{J}$) is induced by an externally-applied field ($\mathbf{E}_{0}$).  We can then express the total electric field within the material in terms of the externally-applied field and the field radiated by the induced current distribution~\cite{Yaghjian1980}
{\begin{align}
\mathbf{E}(\boldsymbol{x}) &=  \mathbf{E}_{0}(\boldsymbol{x}) - \frac{1}{i \omega}\int _V \boldsymbol{\bar{\mathcal{G}}}(\boldsymbol{x},\boldsymbol{x}')\mathbf{J}(\boldsymbol{x}')\;{\mathrm{d}^3x'}\:, \label{eq:original}
\end{align}}
 the volume $V$ is all space and $\boldsymbol{\bar{\mathcal{G}}}$ is a generalized dyadic Green's function, written in terms of the source dyadic ($\mathbf{\bar {L}}$) and the free space Green's function
\begin{align}
\boldsymbol{\bar{\mathcal{G}}}(\boldsymbol{x},\boldsymbol{x}')=\omega^2 \mu_0 \mathop{\mathrm{P.V.}}\left [\mathbf{\bar{I}}+ \frac{1}{k^2} \boldsymbol{\nabla \nabla}\right]  &\frac{e^{i k \left| \boldsymbol{x}-\boldsymbol{x}'\right|}}{4 \pi  \left| \boldsymbol{x}-\boldsymbol{x}'\right|}- \mathbf{\bar{L}} \frac{\delta (\boldsymbol{x}-\boldsymbol{x}')}{ \epsilon_0}\:.
\end{align}
where $\omega$ is the angular frequency of light and P.V. implies a principle value exclusion of $\boldsymbol{x}=\boldsymbol{x}'$ when performing the integration in Eq.~\ref{eq:original}.
 For ease of notation, we can also define a tensor permittivity ($\boldsymbol{\bar{\epsilon}}$) in terms of the conductivity ($\boldsymbol{\bar{\sigma}}$) and electric susceptibility ($\boldsymbol{\bar{\chi}}$)  to relate the induced current to the total electric field
\begin{align}
\boldsymbol{\bar{\epsilon}}\equiv ( \boldsymbol{\bar{\chi}}+1)\epsilon_0 - \frac{ \boldsymbol{\bar{\sigma}}} {i \omega}  \quad \Rightarrow \quad \mathbf{J} (\boldsymbol{x})= - i \omega  \left [ \boldsymbol{\bar{\epsilon}}(\boldsymbol{x})- \epsilon_0 \right ] \mathbf{E}(\boldsymbol{x})\:. \label{eq:induced current}
\end{align}
As we are only considering $\boldsymbol{x}$ in the volume ($V_s$) with non-background permittivity, we can rewrite Eq.~\ref{eq:original} as an integral equation for the induced current
{\begin{align}
i \omega \mathbf{E}_{0}(\boldsymbol{x})   =&- [ {\boldsymbol{\bar {\epsilon}}(\boldsymbol{x}) - \epsilon_0}]^{-1}{\mathbf{J}(\boldsymbol{x})}  +\int _{V_s} \boldsymbol{\bar{\mathcal{G}}}(\boldsymbol{x},\boldsymbol{x}')\mathbf{J}(\boldsymbol{x}') \;{\mathrm{d}^3x'}\:.  \label{eq:integral equation}
\end{align}}
{This integral equation is applicable to the fields of any arbitrary structure in the absence of magnetization.}
  Now, for our purposes, Eq.~\ref{eq:integral equation} has an associated eigenmode equation, which has solutions ${\boldsymbol{v}_i}$ that satisfy
{\begin{align}
i \omega   \lambda_i {\boldsymbol{v}_i}(\boldsymbol{x})    =&  -[ {\boldsymbol{\bar {\epsilon}}(\boldsymbol{x}) - \epsilon_0}]^{-1}{{\boldsymbol{v}_i}(\boldsymbol{x})} +\int _{V_s}  \boldsymbol{\bar{\mathcal{G}}}(\boldsymbol{x},\boldsymbol{x}'){\boldsymbol{v}_i}(\boldsymbol{x}') \;{\mathrm{d}^3x'}\:.   \label{eq:eigenmode equation}
\end{align}}
These eigenmodes can be calculated numerically for simple systems based on discrete dipole approximation~\cite{HopkinsPoddubny2013}, but also for arbitrary continuous structures~\cite{Powell2014}.
Now we can apply symmetry analysis to the structure under consideration.  Each irreducible representation of the structure's highest-order symmetry group will describe the transformation properties of a distinct set of eigenmodes.  In the absence of accidental degeneracies, each of these eigenmodes will have a unique eigenvalue with a degeneracy level equal to the dimension of the associated irreducible representation \cite{Dresselhaus2008}.
For our purposes we are considering the $C_n$, $S_n$, $D_n$ and $C_{nh}$ symmetry groups, and, specifically, the two-dimensional $E$ irreducible representations, which describe the transformation properties of any normally-incident planewave ($\mathbf{E}_{0}$ in Eq.~\ref{eq:integral equation}) under the given group's symmetry operations.   Because of this, only the eigenmodes which transform according to these $E$ irreducible representations can be excited.
It is now important to notice here that we are working in a complex space. It implies that the two-dimensional $E$ irreducible representations of the $C_n$, $S_n$ and $C_{nh}$ groups are written as two, one-dimensional irreducible representations that are complex conjugates of each other (for example, see Table~\ref{character table}). In regard to notation, we will refer to these two, one-dimensional irreducible representations as $E^{+}$ and $E^{-}$ (i.e. where ${E^{+}}^* = E^{-}$).
{ \begin{table} [h!]
\centering
\begin{tabular}{c " c  c   c  } 
\thickhline  \\[-2.5ex]
$ \qquad$ { \it \large C$_3$} $\qquad$ & $\qquad$ { \it \large E}  & $\qquad$ { \it \large C$_3$}   & $\qquad$ { \it \large C$_{3}^{\;2}$} \\ [0.5ex] 
\thickhline 
 \\[-2.5ex]
$\qquad$ { \it \large A} $\qquad$ & $\quad \qquad1$ &  $\quad \quad \; 1$ & $\quad \quad \;1$  \\ [2ex] 
$\quad \;$ { \it \large E}$\qquad \;$ & $\qquad \; \left \{ \begin{array}{c} 1 \\ 1\end{array}
\right.$ &  $\qquad \; \begin{array}{c} \phi  \\ \phi^2 \end{array} $ &  $ \qquad \; \begin{array}{c} \phi^2 \\ \phi  \end{array}$ \\ [4ex] 
\thickhline
\end{tabular}
{\large\begin{equation*} \huge
\phi = e^{2 \pi i / 3}
\end{equation*}}\\[-1ex] 
\caption{The character table of the 3-fold rotational ($C_3$) symmetry group.  Note here,  the  two-dimensional $E$ representation is made up of two, one-dimensional irreducible representations that are complex conjugates of each other.  }
\label{character table}
\end{table}}

Taking into account that $E^{+}$ and $E^{-}$ are different irreducible representations, each will describe the transformation properties of a distinct set of eigenmodes, which we  refer here to as $\{{\boldsymbol{v}^{\mbox{\footnotesize \bf   +}}}\}$ and $\{{\boldsymbol{v}^{\mbox{\footnotesize \bf   -}}}\}$, respectively.
From the geometrical symmetry perspective, the eigenmodes associated with different irreducible representations should not be degenerate~\cite{Dresselhaus2008}.
 However, such eigenmodes can become degenerate in some systems where additional symmetries are present.
One of the key symmetries of electromagnetic theory that is neglected in a purely geometric argument is Lorentz reciprocity\footnote{Lorentz reciprocity itself can be seen as an application of the general Onsager reciprocity principle to electromagnetics, which in turn follows from the  behavior of dissipative equilibrium systems under the time reversal operation~\cite{landau9}. However, for the discussion here, it is more convenient to use Lorentz reciprocity argument directly.}.
In the following we will prove  that every eigenmode in $E^+$ will always be degenerate with one eigenmode of the associated $E^-$ (and {\it vice versa}), purely due to the inherent Lorentz reciprocity of the Maxwell's equations.

We start by taking an eigenmode (${\boldsymbol{v^{\mbox{\tiny \bf +}}}_i}$), which transforms according to a one-dimensional $E^+$ irreducible representation.
 We also impose the standard normalization condition
{\begin{align}
\int_{V_s}{\boldsymbol{v^{\mbox{\tiny \bf +}}}_i}^* \cdot {\boldsymbol{v^{\mbox{\tiny \bf +}}}_i}\;\mathrm{d}V = 1\:, \label{eq:start}
\end{align}}
where the dot denotes a vector dot product.   Importantly, if ${\boldsymbol{v^{\mbox{\tiny \bf +}}}_i}$ transforms under symmetry operations according to $E^+$, we know that ${\boldsymbol{v^{\mbox{\tiny \bf +}}}_i}^*$ must transform according to the complex conjugate irreducible representation, $E^-$, because any operator describing a geometric symmetry operation in a Euclidean space will be real.
We can therefore write ${\boldsymbol{v^{\mbox{\tiny \bf +}}}_i}^*$ as some linear combination of the eigenmodes, $\{{\boldsymbol{v}^{\mbox{\footnotesize \bf   -}}}\}$, which transform according to the  $E^-$ irreducible representation
{\begin{align}
{\boldsymbol{v^{\mbox{\tiny \bf +}}}_i}^* = \sum \limits_j b_{ij} {\boldsymbol{v^{\mbox{\footnotesize \bf   -}}}_j}  \qquad b_{ij} \in \mathbb{C}\:. \label{eq:basis}
\end{align}}
By  substituting Eq.~\ref{eq:basis}  into Eq.~\ref{eq:start} we can rewrite the normalization of ${\boldsymbol{v^{\mbox{\tiny \bf +}}}_i}$ as
{\begin{align}
\sum \limits_j b_{ij}  \left( \int_{V_s} {\boldsymbol{v^{\mbox{\footnotesize \bf   -}}}_j} \cdot {\boldsymbol{v^{\mbox{\tiny \bf +}}}_i}\;\mathrm{d}V\right) =1\:.  \label{eq:substitution}
\end{align}}
 For Eq.~\ref{eq:substitution} to hold we know that there must be at least one $j$ such that
{\begin{align}
\int _{V_s} {\boldsymbol{v^{\mbox{\footnotesize \bf   -}}}_j} \cdot {\boldsymbol{v^{\mbox{\tiny \bf +}}}_i} \;\mathrm{d}V \neq 0\:. \label{eq:requirement}
\end{align}}
We now consider the role of the Lorentz reciprocity, a consequence of which is that both the dyadic Greens function and the permittivity tensor must be symmetric, albeit complex and not necessarily Hermitian
{\begin{align}
\boldsymbol{\bar{\mathcal{G}}}(\boldsymbol{x},\boldsymbol{x}')  = \boldsymbol{\bar{\mathcal{G}}}(\boldsymbol{x}',\boldsymbol{x}),  \quad\boldsymbol{\bar{\mathcal{G}}}=  \boldsymbol{\bar{\mathcal{G}}}^{T},\quad {\boldsymbol{\bar {\epsilon}}} = {\boldsymbol{\bar {\epsilon}}}^{T}\;.
\end{align}}
Due to this symmetry, it is possible to write the overall operator of the eigenvalue equation (Eq.~\ref{eq:eigenmode equation}) as a matrix in the normal form shown by Gantmacher~\cite{Gantmacher1959}.  A result of this is that nondegenerate eigenmodes must be orthogonal under unconjugated projections~\cite{Craven1969}
{\begin{align}
\int_{Vs}  {\boldsymbol{v}_\alpha} \cdot {\boldsymbol{v}_\beta}   \;\mathrm{d}V = 0 \quad \mathrm{when}\;\,\lambda_\alpha \neq \lambda_\beta\:. \label{eq:this is responsible for orthogonality}
\end{align}}
Hence Eq.~\ref{eq:requirement} can be true only for some ${\boldsymbol{v^{\mbox{\tiny \bf +}}}_i}$ is if there exists an eigenmode ${\boldsymbol{v^{\mbox{\footnotesize \bf   -}}}_j}$ that is degenerate with ${\boldsymbol{v^{\mbox{\tiny \bf +}}}_i} $.
Moreover, we must always be able to find at least one  ${\boldsymbol{v^{\mbox{\footnotesize \bf   -}}}_j}$   that satisfies Eq.~\ref{eq:requirement} for {\it each} ${\boldsymbol{v^{\mbox{\tiny \bf +}}}_i}$.
However, suppose now that multiple ${\boldsymbol{v^{\mbox{\footnotesize \bf   -}}}_k}$ satisfied Eq.~\ref{eq:requirement} for the one ${\boldsymbol{v^{\mbox{\tiny \bf +}}}_i} $; this would mean that each of these ${\boldsymbol{v^{\mbox{\footnotesize \bf   -}}}_k}$ are degenerate with every other  ${\boldsymbol{v^{\mbox{\footnotesize \bf   -}}}_k}$.  
Such unenforced degeneracies between eigenmodes are accidental degeneracies and can be excluded from the general consideration.  
We can therefore expect each ${\boldsymbol{v^{\mbox{\footnotesize \bf   -}}}_j}$ to be degenerate with a single ${\boldsymbol{v^{\mbox{\tiny \bf +}}}_i} $ (and {\it vice versa}).  
For ease of notation, we will use the subscript convention that ${\boldsymbol{v^{\mbox{\tiny \bf +}}}_i} $ is degenerate with ${\boldsymbol{v^{\mbox{\footnotesize -}}}_i} $. 
We can now combine Eqs.~\ref{eq:start} and \ref{eq:basis} to write the following relation
{\begin{align}
b_{ij}\int_{V_s} {\boldsymbol{v^{\mbox{\tiny \bf +}}}_i} \cdot {\boldsymbol{v^{\mbox{\footnotesize \bf   -}}}_j}  \;\mathrm{d}V &=  \delta_{ij}\:,   \label{eq:delta}
\end{align}}
where $\delta_{ij}$ is a Kronecker delta function.
This completes the proof, we have shown that Lorentz reciprocity forces degeneracy between pairs of eigenmodes of complex conjugate irreducible representations in {\it any} symmetry group.
For our purposes, we will specifically identify the $E$ irreducible representations of $C_n$, $S_n$ and $C_{nh}$ symmetry groups ($n \geq 3$) as being subject to this argument (the degeneracy already exists in $D_n$ from geometry alone).   This result is applicable to all associated symmetric scattering systems irrespective of their specific dimensions or constituent materials.

To emphasize the significance of this degeneracy, we have to relate the eigenmodes to the applied fields that excite them.
The behavior of the $E^+$ and $E^-$ irreducible representations in $C_n$, $S_n$ and $C_{nh}$ symmetry groups is that symmetric rotations are described by uniform phase shifts that are complex conjugates of each other.  At normal incidence, it is the  RCP and LCP plane    waves that behave in this way and they can be, therefore, assigned to $E^+$ and $E^-$, respectively.
As such, we can define the eigenmodes $\{{\boldsymbol{v^{\mbox{\tiny \bf +}}}_i} \}$ to correspond to those excited by an RCP plane wave and  $\{{\boldsymbol{v}^{\mbox{\footnotesize \bf   -}}}\}$ to those excitable by an LCP plane wave.
Therefore the degeneracy we have just proven means that the modes that can be excited by LCP and RCP plane waves are degenerate.  
As such, changing the polarization of the incident plane wave between LCP and RCP  cannot result in the excitation of modes that were previously forbidden by symmetry (dark modes).  
Moreover, the degeneracy of left- and right-handed eigenmodes highlights an important conceptual point on why the handedness of the incident field is able to affect optical response.  
While LCP and RCP plane waves are enantiomers, the degenerate eigenmodes they respectively excite cannot themselves be enantiomers because the geometry is not conserved under reflections.    
This permits nontrivial differences to exist in the distribution of eigenmodes that are excited by LCP and RCP plane waves. 
In other words, circular dichroism and other chiral scattering effects will be due to the varying magnitudes and phases of the excitations of degenerate eigenmodes.
To now derive physically meaningful information from the current distributions, we need to calculate the extinction and absorption cross-sections to define circular dichroism.
 These two cross-sections can be calculated from current distributions similar to point-dipole systems in Ref.~\cite{Markel1995} by defining an infinite number of point dipoles $\{\mathbf{p}\}$, where each dipole moment is defined for some infinitesimal volume $\mathrm{d}V$ over which $ \mathbf{J}$ can be considered as constant
{\begin{align}
\mathbf{p}_{\boldsymbol{x}} = -\frac{1}{i \omega} \int_{\mathrm{d}V}  \mathbf{J} (\boldsymbol{x}')  {\mathrm{d}^3 x'} =     -\frac{1}{i \omega} \mathbf{J} (\boldsymbol{x}) \mathrm{d}V\:.
\end{align}}
Using the definition for induced current in Eq.~\ref{eq:induced current} we can further define the polarizability of these dipoles as
{\begin{align}
 \boldsymbol{\bar{\alpha}}_{\boldsymbol{x}} =   \left [ \boldsymbol{\bar{\epsilon}}(\boldsymbol{x})- \epsilon_0 \right ] \mathrm{d}V\:.
\end{align}}
The extinction can then be determined from the projection of the complex conjugated incident field onto the induced current
{\begin{align}
\sigma_{\mathrm{ext}} &= \frac{k}{\epsilon_0 |\mathbf{E}_0|^2}\; \mathrm{Im}\left \{ \sum \limits_{\boldsymbol{x}} \mathbf{E}_0^* \cdot \mathbf{p}_{\boldsymbol{x}}  \right \} \nonumber \\ &=  \frac{1}{|\mathbf{E}_0|^2}\sqrt{\frac{\mu_0}{\epsilon_0}} \; \mathrm{Re}\left \{ \int_{V_s} \mathbf{E}_0^* \cdot \mathbf{J}  \; \mathrm{d}V \right \}\:. \label{eq:extinction definition}
\end{align}}
Similarly, we can express the absorption in terms of the intensity of the induced currents by keeping only the lowest order of $\mathrm{d}V$
{\begin{align}
\sigma_{\mathrm{abs}} &=-\frac{k}{{\epsilon_0}^2 |\mathbf{E}_0|^2} \sum \limits_{x}  \mathbf{p}_{\boldsymbol{x}}^*  \cdot \left( \mathrm{Im} \left \{ \boldsymbol{\bar{\alpha}}_{\boldsymbol{x}}^{-1} \right \}  +\frac{k^3}{6 \pi} \right) \;  \cdot \mathbf{p}_{\boldsymbol{x}} \nonumber\\
&=-\frac{1}{\omega c {\epsilon_0}^2 |\mathbf{E}_0|^2}  \int_{V_s}\mathbf{J}^*  \cdot\mathrm{Im} \left \{\left ( \boldsymbol{\bar{\epsilon}}(\boldsymbol{x})- \epsilon_0 \right )^{-1} \right \}  \cdot \mathbf{J} \;\mathrm{d}V\:. \label{eq:absorption definition}
\end{align}}
Hence, it is straightforward now to calculate the circular dichroism in extinction and absorption cross-sections from the currents that are induced by circularly-polarized plane waves with opposite handedness.
Importantly, due to the optical theorem, the extinction cross-section depends only on the forward scattering amplitude whereas the  scattering and absorption cross-sections depend on the far-field scattering at all angles (or, equivalently, on the full structure of the near-field).
This already suggests that the scattering and absorption cross-sections will be more sensitive to the polarization in the structures with reduced symmetry, a hypothesis which we will confirm in the coming arguments.

\section{Planar chiral oligomers}

Circular dichroism is traditionally defined as a difference in optical loss between LCP and RCP plane waves propagating in the same direction. However, given the freedom of two polarizations and two propagation directions, there are {\it four} distinct excitations that can be applied by circularly-polarized plane waves. 
These four circularly-polarized plane waves be can represented in terms of their polarization depicted spatially along the propagation direction, where each of them resembles either one of two oppositely-handed helices (seen in Fig.~\ref{fig:helices}).  The remaining distinction is then the direction such a `polarization helix' rotates in time.
 This depiction of an incident field corresponds to the spatial distribution of the applied electric field, which is what we use in our integral equation approach for the induced currents (Eq.~\ref{eq:integral equation}).
{ \begin{figure*}[t!]
\centering
\centerline{\includegraphics[width=\textwidth]{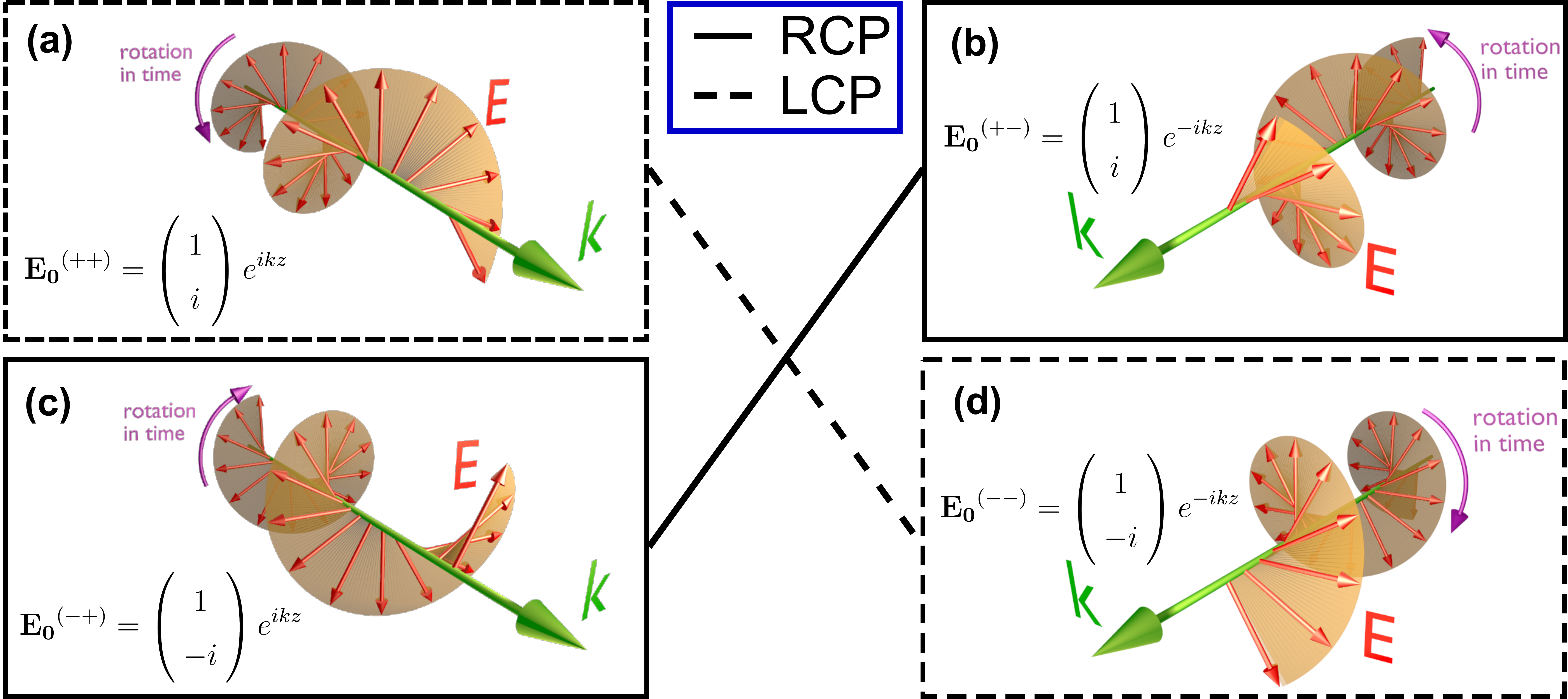}}
\caption{Schematic illustration of the electric field excitations that can be exerted by a circularly-polarized plane wave given four different combinations of propagation direction and polarization sign.
}
\label{fig:helices}
\end{figure*}}
In the expressions for the applied electric field, seen as insets in {Fig.~\ref{fig:helices}(a)-(d)}, we use the notation of two superscript plus and/or minus signs to indicate the sign of the imaginary unit in the polarization vector and exponential, respectively.  This notation highlights which plane waves are complex conjugates (reciprocal), {\it e.g.} $\mathbf{E_0}^{(\pm \mp)}$ is the complex conjugate of $\mathbf{E_0}^{(\mp \pm)}$.

The reason circular dichroism is defined for one propagation direction only ({\it i.e.} two, not four, excitations)  is reciprocity, which equates the extinction cross-section of oppositely propagating plane waves and, thus, the difference in extinction cross-section between LCP and RCP plane waves does not depend on the propagation direction~\cite{Bohren1983}.
This invariance between reciprocal plane waves can be observed in the electric field distributions ($\mathbf{E}_{0}(\boldsymbol{x})$) that are complex conjugates of each other.
So, if we then start from the earlier analysis of the induced currents in structures with $C_n$, $D_n$, $S_n$ or $C_{nh}$ symmetry, we are able to express the current ($\mathbf{J} $) induced by a circularly-polarized plane wave ($\mathbf{E}_{0}$) in terms of eigenmodes
{\begin{align}
\mathbf{E}_{0} = \sum \limits _i a_i \lambda_i {\boldsymbol{v^{\mbox{\tiny \bf +}}}_i}  \quad \Rightarrow \quad \mathbf{J} = \sum \limits _i a_i {\boldsymbol{v^{\mbox{\tiny \bf +}}}_i}\:. \label{eq:J definition}
\end{align}}
Using Eq.~\ref{eq:basis}, we can also define the current ($\mathbf{J'} $) induced by the complex conjugate field ($\mathbf{E}_{0}^*$)
{\begin{align}
\mathbf{E}_{0}^* &= \sum \limits _{i,j} a_i^* \lambda_i ^* b_{ij} {\boldsymbol{v^{\mbox{\footnotesize \bf   -}}}_j} \quad \Rightarrow \quad \mathbf{J'} = \sum \limits  _{i,j} a_i^* b_{ij}\frac{\lambda_i^*} {\lambda_j} {\boldsymbol{v^{\mbox{\footnotesize \bf   -}}}_j}\:.  \label{eq:J' definition}
\end{align}}
This shows that the overall excitation of each degenerate eigenmode pair is not necessarily the same between $\mathbf{E}_{0}$ and $\mathbf{E}_{0}^*$.  However, despite the change in eigenmode excitation, we can use the result in Eq.~\ref{eq:delta} with Eqs.~\ref{eq:J definition} and \ref{eq:J' definition} to equate the extinction (Eq.~\ref{eq:extinction definition}) of complex conjugate applied fields (as expected)
{\begin{align}
\int_{V_s} \mathbf{E}_{0}^* \cdot \mathbf{J} \;\mathrm{d}V = \int_{V_s} \mathbf{E}_{0} \cdot \mathbf{J'} \;\mathrm{d}V\:. \label{eq:reciprocity}
\end{align}}
This, therefore, demonstrates that the total excitation of each nondegenerate eigenmode can change between reciprocal plane-waves while the extinction is conserved.
The conservation of extinction means a structure with a planar reflection symmetry (such as in $C_{nh}$ symmetry) cannot exhibit circular dichroism in extinction.
Rigorously, applying a symmetric reflection operator ($\hat{\sigma}_h$) to the global reference frame of the generic system in Eq.~\ref{eq:J definition} will not change the scatterer's geometry by definition, but the applied field and the induced current become $\hat{\sigma}_h  \mathbf{E}_{0} $  and $\hat{\sigma}_h \mathbf{J}$ when expressed in terms of the new reference frame.
Given that it is a unitary operation, the $\hat{\sigma}_h$ operator then cancels when evaluating the extinction for $\hat{\sigma}_h  \mathbf{E}_{0} $ and $\hat{\sigma}_h \mathbf{J}$, leaving the extinction unchanged
{\begin{align}
\int_{V_s} \left( \hat{\sigma}_h \mathbf{E}_{0}\right)^* \cdot \left( \hat{\sigma}_h \mathbf{J}   \right) \;\mathrm{d}V
&=  \int_{V_s} \mathbf{E}_{0}^*\cdot \mathbf{J}\;\mathrm{d}V\:.  \label{eq:unitary}
\end{align}}
As the reflection operator changes the propagation direction of the incident field  while leaving the polarization unchanged, $\hat{\sigma}_h  \mathbf{E}_{0} $ corresponds to an oppositely-handed polarization helix to that of $\mathbf{E}_{0} $ ({\it cf.} Fig.~\ref{fig:helices}(a) and (b), and Fig.~\ref{fig:helices}(c) and (d)).
 Therefore Eq.~\ref{eq:unitary}, in conjunction with Eq.~\ref{eq:reciprocity}, means that all four excitations from a circularly-polarized plane-wave will produce the same extinction cross-section and, subsequently, this shows that no circular dichroism can occur in extinction for structures with $C_{nh}$ symmetry.
Our argument also holds for structures with $S_n$ symmetry if we substitute the $\hat{\sigma}_h$ reflection operation for the $\hat{S}_{n}$ improper rotation operation.
As such, we have actually shown that neither $C_{nh}$ or $S_n$ symmetries permit circular dichroism in extinction.
More specifically, we demonstrated that this conserved extinction does not require there to be a trivial difference between the currents induced by reciprocal plane-waves.

Up to this point, the direction that an applied field rotates in time has been discounted because of the reciprocity condition in Eq.~\ref{eq:reciprocity}.
However it is worth acknowledging that this result is only applicable to extinction and, subsequently, in the presence of material losses, reciprocity does not constrain the absorption and scattering cross-sections independently. 
Therefore, by  introducing material losses to produce an absorption cross-section, neither the absorption or scattering cross-section are necessarily invariant under reciprocal plane-waves.
This is of particular interest for planar chiral scattering geometries, such as those with $C_{nh}$ symmetry, which we have shown cannot produce circular dichroism in the extinction cross-section due to the combination of reciprocity and planar reflection symmetry.
{Indeed, the combination of polarization independent extinction and varying absorption requires that a resonance in absorption for one circular polarization must coincide with a suppression in the scattering cross-section to balance the conserved extinction.}

To consider the origin of such an effect, we will begin by proving that the total induced current intensity can only change between reciprocal excitations if the structure's eigenmodes are nonorthogonal.
Specifically, if we assume that the scattering structure's eigenmodes are orthogonal to each other, we can define the excitations of each eigenmode by reciprocal excitations explicitly using complex projections
{\begin{align}
&\mathbf{E}_{0} = \sum \limits _i \left( \int_{V_s} {\boldsymbol{v^{\mbox{\tiny \bf +}}}_i}^* \cdot \mathbf{E}_{0}   \;\mathrm{d}V \right) {\boldsymbol{v^{\mbox{\tiny \bf +}}}_i}\nonumber \\
&\quad \Rightarrow  \quad \mathbf{J} = \sum \limits _i \left( \int_{V_s} {\boldsymbol{v^{\mbox{\tiny \bf +}}}_i}^* \cdot \mathbf{E}_{0}   \;\mathrm{d}V \right)\frac{{\boldsymbol{v^{\mbox{\tiny \bf +}}}_i}}{\lambda_i} \label{eq:one}\:, \\
&\mathbf{E}_{0}^* = \sum \limits _i \left( \int_{V_s} {\boldsymbol{v}^{\mbox{\footnotesize \bf   -}}}_{i}^* \cdot \mathbf{E}_{0}^*   \;\mathrm{d}V \right) {\boldsymbol{v}^{\mbox{\footnotesize \bf   -}}}_{i} \nonumber\\
&\quad \Rightarrow \quad \mathbf{J'} = \sum \limits _i \left( \int_{V_s} {\boldsymbol{v}^{\mbox{\footnotesize \bf   -}}}_{i}^* \cdot \mathbf{E}_{0}^*   \;\mathrm{d}V \right)\frac{ {\boldsymbol{v}^{\mbox{\footnotesize \bf   -}}}_{i}}{\lambda_i}\:.\label{eq:two}
\end{align}}
Similarly we can also define $b_{ij}$ in Eq.~\ref{eq:basis} as
{\begin{align}
b_{ij}= \int_{V_s} {\boldsymbol{v^{\mbox{\footnotesize \bf   -}}}_j}^* \cdot {\boldsymbol{v^{\mbox{\tiny \bf +}}}_i}^* \;\mathrm{d}V =  \left(\int_{V_s} {\boldsymbol{v^{\mbox{\footnotesize \bf   -}}}_j} \cdot {\boldsymbol{v^{\mbox{\tiny \bf +}}}_i} \;\mathrm{d}V \right)^*\:.  \label{eq:yay}
\end{align}}
Notably, from Eq.~\ref{eq:this is responsible for orthogonality}, this means $b_{ij}$ is zero if $i \neq j$.  Then we can relate the excitations of each eigenmode, under complex conjugate incident fields, by substituting Eq.~\ref{eq:yay} into Eq.~\ref{eq:basis}
{\begin{align}
 \left( \int_{V_s} {\boldsymbol{v^{\mbox{\tiny \bf +}}}_i}^* \cdot \mathbf{E}_{0}   \;\mathrm{d}V \right)^* =  \left(\int_{V_s} {\boldsymbol{v^{\mbox{\tiny \bf +}}}_i} \cdot {\boldsymbol{v}^{\mbox{\footnotesize \bf   -}}}_{i}   \;\mathrm{d}V \right)  \int_{V_s} {\boldsymbol{v}^{\mbox{\footnotesize \bf   -}}}_{i}^* \cdot \mathbf{E}_{0}^*  \;\mathrm{d}V  \:. \label{eq:A}
\end{align}}
At the same time, Eq.~\ref{eq:yay}, in conjunction with Eq.~\ref{eq:delta}, means that the real projection of one eigenmode onto its degenerate partner has unit magnitude
{\begin{align}
 \left( \int_{V_s} {\boldsymbol{v}^{\mbox{\footnotesize \bf   -}}}_{i}  \cdot {\boldsymbol{v^{\mbox{\tiny \bf +}}}_i}  \;\mathrm{d}V \right)^{-1} =  \left( \int_{V_s} {\boldsymbol{v}^{\mbox{\footnotesize \bf   -}}}_{i} \cdot {\boldsymbol{v^{\mbox{\tiny \bf +}}}_i}  \;\mathrm{d}V \right)^*\:. \label{eq:B}
\end{align}}
To combine this all, we start from the definitions of  $\mathbf{J}$ and $\mathbf{J'}$ in Eqs.~\ref{eq:one} and \ref{eq:two}, then use the relations in Eqs.~\ref{eq:A} and \ref{eq:B}, in addition to our original assumption that all eigenmodes are orthogonal, to get the result that
{\begin{align}
\int_{V_s} \mathbf{J'}^* \cdot \mathbf{J'} \;\mathrm{d}V  = \int_{V_s} \mathbf{J}^* \cdot \mathbf{J}\;\mathrm{d}V\:. \label{eq:isotropic result}
\end{align}}
 For geometries made of isotropic materials, this means that circular dichroism in absorption (Eq.~\ref{eq:absorption definition}) ceases to exist between complex conjugate excitations.   
In other words, we have proven that absorption circular dichroism can only exist between reciprocal plane-waves if such geometries have nonorthogonal eigenmodes.
{ This peculiar requirement means circular dichroism in absorption can be considered as an interference effect in the same sense that Fano resonances are.
Specifically, in Ref.~\cite{HopkinsPoddubny2013}, it was shown that the nonzero projections of one eigenmode onto another eigenmode are able to fully describe Fano resonances.   
We can now establish a link between circular dichroism in absorption and this eigenmode overlap by relating it to the $b_{ij}$ coefficient using Eqs.~\ref{eq:basis} and \ref{eq:delta}}
\begin{align}
\int {\boldsymbol{v^{\mbox{\tiny \bf +}}}_i}^{*} \cdot {\boldsymbol{v^{\mbox{\tiny \bf +}}}_j} \;\mathrm{d}V = b_{ij} \int {\boldsymbol{v^{\mbox{\footnotesize \bf   -}}}_j} \cdot {\boldsymbol{v^{\mbox{\tiny \bf +}}}_j}\;\mathrm{d}V \:.
\end{align}
{ This shows us that $b_{ij}$ is proportional to the overlap between the nondegenerate eigenmodes $ {\boldsymbol{v^{\mbox{\tiny \bf +}}}_i}$ and  $ {\boldsymbol{v^{\mbox{\tiny \bf +}}}_j}$.
 If we then refer to Eqs.~\ref{eq:J definition} and \ref{eq:J' definition}, we can see that the excitation of nondegenerate eigenmodes should vary more between reciprocal excitations given the presence of a large $b_{ij}$.
In other words, we should observe significant circular dichroism in absorption and scattering cross-sections when there is a large eigenmode overlap.  
Moreover, a large eigenmode overlap is known to exist, quite prominently, at a Fano resonance. 
So, to produce circular dichroism in absorption, we can take a structure known to produce Fano resonances and alter it to be planar chiral.}
In Fig.~\ref{fig:chiral oligomer} we begin with a gold heptamer which is known to support Fano resonances~\cite{Hentschel2010} and then alter the nanoparticles in the outer ring to make it both planar chiral and having $C_{nh}$ symmetry.   The choice of parameters was chosen so that the central particle resonance is overlapped with that of the outer ring~\cite{Miroshnichenko2012}.
{\begin{figure*}[ht!]
\centerline{\includegraphics[width=0.75\textwidth]{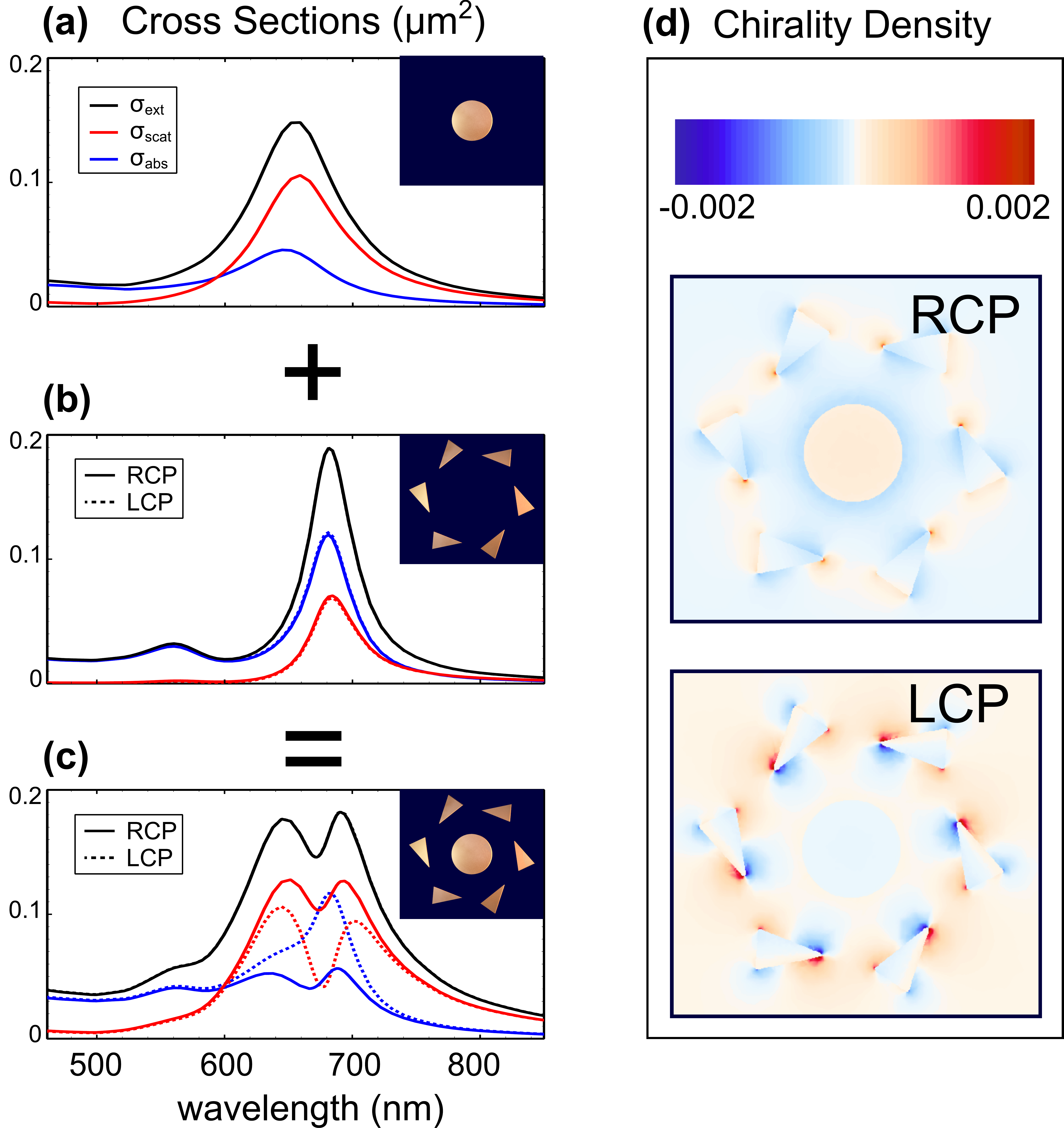}}
\caption{
Simulations demonstrating the role of interference for inducing circular dichroism in the absorption cross-section of a planar chiral heptamer with $C_{6h}$ symmetry.
We observe the creation of significant circular dichroism in absorption in the vicinity of the Fano resonance, as is predicted given it corresponds to high modal overlap.  Additionally, on the right hand side, we show that this circular dichroism can also be observed in the magnitude of the near-field chirality density\cite{Bliokh2011} at the Fano resonance.  Here the chirality density induced by an RCP plane-wave is less than half of that for the LCP case.
The calculation parameters are given in text.}
\label{fig:chiral oligomer}
\end{figure*}}
As the outer ring in Fig~\ref{fig:chiral oligomer} is a planar chiral structure, it will experience different current distributions in response to LCP and RCP plane waves (see Eqs.~\ref{eq:J definition} and \ref{eq:J' definition}), but in isolation it does not experience significant circular dichroism. However the collective structure exhibits a Fano resonance, which leads to significant circular dichroism in absorption and scattering. The extent of the circular dichroism is, in fact, sufficient to swap the dominance of scattering and absorption cross-sections using polarization.
This therefore supports our derivation that nonreciprocal circular dichroism in absorption is an interference effect.

In Fig.~\ref{fig:chiral oligomer}, all nanoparticles are made from 20~nm thick gold, the central disk has a diameter of 140nm and each triangular nanoparticle has major and minor axes of 100~nm and 60~nm (respectively).   The triangular nanoparticles have been placed at a radius of 170~nm away from the center of the disk and the major axis is oriented 65\degree~off the radial vector.  LCP and RCP are defined relative to a vector pointing out of the page.  All simulations were performed using CST Microwave Studio and gold data was taken from Johnson and Christy~\cite{goldData}.
We also demonstrate  that the effect is robust and is not dependent on precise parameters.  In Fig.~\ref{fig:parameter sweep}(a) and (b), we plot the difference between LCP and RCP absorption cross-sections when varying both the diameter of the outer ring of triangular nanoparticles and their angular orientation.   It can be seen that the significant splitting between absorption from LCP and RCP plane waves, observed first in Fig.~\ref{fig:chiral oligomer}, is evident for a wide range of particle arrangements.  Furthermore, in Fig.~\ref{fig:parameter sweep}(c), we show that the same effect will occur even when changing the number of constituent nanoparticles that make up the oligomer.  This subsequently shows that circular dichroism in absorption is a robust and widely achievable feature of geometries that produce modal interference.
{\begin{figure*}[h!]
\centerline{\includegraphics[width=\textwidth]{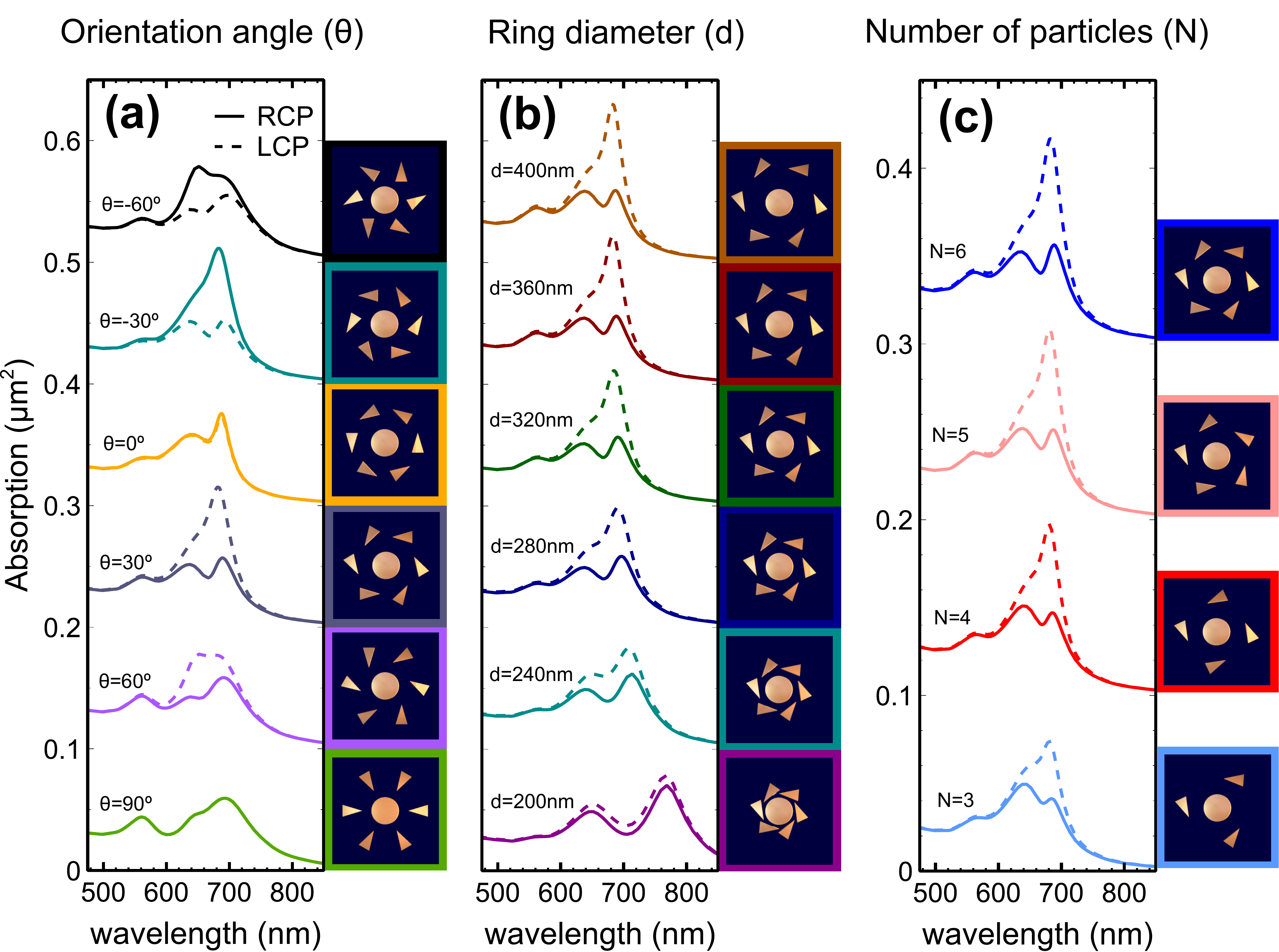}}
\caption{
Simulations of the absorption cross-section produced by LCP and RCP plane waves incident on the same planar chiral oligomer seen in Fig.~\ref{fig:chiral oligomer}, when: (a) rotating the triangular nanoparticles in 30\degree~increments, (b) varying the diameter of the ring of triangular nanoparticle in 40nm increments and (c) reducing the number of nanoparticles.  It can be seen that circular dichroism in absorption is most dependent on the orientation of the triangular nanoparticles, but the effect nonetheless exists for a wide range of structures.  All dimensions are the same as Fig.~\ref{fig:chiral oligomer}, except where mentioned otherwise. }
\label{fig:parameter sweep}
\end{figure*}}

To investigate the associated physical dependencies of circular dichroism in the absorption cross-section, we need to identify a physical characteristic that causes nonorthogonal eigenmodes.
This is in fact quite straightforward if we neglect the retardation of coupling within the structure, which can be done either resorting to the quasistatic approximation ($k\to0$) or by working in the near-field limit ($ k | \boldsymbol{x} - \boldsymbol{x}'|\to 0$) for small systems. {
 In these cases, the free space Green's function becomes entirely real.
So we can take the complex conjugate of the eigenmode equation of our system (Eq.~\ref{eq:eigenmode equation}) to get the following equation}
{\begin{align}
i \omega \eta_i{\boldsymbol{v^{\mbox{\tiny \bf +}}}_i}^*(\boldsymbol{x})  =  &-( {\boldsymbol{\bar {\epsilon}}(\boldsymbol{x}) - \epsilon_0})^{-1}{{\boldsymbol{v^{\mbox{\tiny \bf +}}}_i}^*(\boldsymbol{x})} +\int _{V_s}  \boldsymbol{\bar{\mathcal{G}}}(\boldsymbol{x},\boldsymbol{x}'){\boldsymbol{v^{\mbox{\tiny \bf +}}}_i}^*(\boldsymbol{x}') \;{\mathrm{d}^3x'} \:,
\end{align}}
where
{\begin{align}
\eta_i =   \frac{1}{i \omega}\left [ {\left( {\boldsymbol{\bar {\epsilon}}\,^*(\boldsymbol{x})- \epsilon_0^*}\right)}^{-1}- \left({\boldsymbol{\bar {\epsilon}} (\boldsymbol{x}) - \epsilon_0}\right)^{-1}  \right ]   -\lambda_i^*\:.
\end{align}}
Notably, the complex conjugate of any eigenmode (${\boldsymbol{v^{\mbox{\tiny \bf +}}}_i}^* $) will be an eigenmode in its own right if there is only one uniform and isotropic material in the structure ({\it i.e.} if ${\boldsymbol{\bar {\epsilon}} (\boldsymbol{x})\rightarrow\epsilon}$).
Then, given ${\boldsymbol{v^{\mbox{\tiny \bf +}}}_i}^*$ is an eigenmode, we can utilize Eqs.~\ref{eq:start} and  \ref{eq:basis} to get the result that
{\begin{align}
b_{ij}=\delta_{ij}  \quad \mathrm{and}\quad   {\boldsymbol{v}^{\mbox{\footnotesize \bf   -}}}_{i} = {\boldsymbol{v^{\mbox{\tiny \bf +}}}_i}^*\:. \label{eq:orthogonality}
\end{align}}
From Eq.~\ref{eq:this is responsible for orthogonality}, this is sufficient to ensure that the eigenmodes are orthogonal.
So, if a structure is made of a single isotropic material, nonorthogonal eigenmodes can only exist if there is retarded coupling within the structure.
This means, for instance, that dichroism between reciprocal, circularly-polarized plane-waves in absorption does not occur for a single-material structure in the quasistatic approximation.
However, if one wants to work in the quasistatic regime and still observe the circular dichroism in absorption, the nonorthogonality of eigenmodes can instead be the result of having anisotropic or inhomogeneous materials ({\it i.e.} even if we neglect retardation).
This can be seen from the fact that any permittivity distribution $\boldsymbol{\bar{\epsilon}}(\boldsymbol{x})$ represents the geometry and therefore has to be invariant under symmetry operations.  So, if we were to define an incident field as
\begin{align}
\mathbf{E'_0}(\boldsymbol{x}) = - \left[ \boldsymbol{\bar{\epsilon}}(\boldsymbol{x})-\epsilon\right]^{-1}{\boldsymbol{v^{\mbox{\tiny \bf +}}}_i}(\boldsymbol{x})\:.
\end{align}
then this incident field would transform under symmetry operations according to ${\boldsymbol{v^{\mbox{\tiny \bf +}}}_i}(\boldsymbol{x})$ and we could subsequently express it as some linear combination  of the eigenmodes $\{\boldsymbol{v}^{\mbox{\tiny \bf +}} \}$

 {\begin{align}
-\left[  \boldsymbol{\bar{\epsilon}}(\boldsymbol{x})-\epsilon\right]^{-1}{\boldsymbol{v^{\mbox{\tiny \bf +}}}_i}(\boldsymbol{x})= \sum \limits _j u_{ij}{\boldsymbol{v^{\mbox{\tiny \bf +}}}_j}(\boldsymbol{x})\:. \label{eq:inhomogeneous}
\end{align}}
Provided that ${\boldsymbol{v^{\mbox{\tiny \bf +}}}_i} $ is not, by chance, an eigenmode of $\left(  \boldsymbol{\bar{\epsilon}}(\boldsymbol{x})-\epsilon\right)^{-1}$, there has to be at least one ${k\neq i}$ such that $u_{ik}$ is nonzero.
Referring back to our original eigenmode equation (Eq.~\ref{eq:eigenmode equation}), this means that the Green's function integral of ${\boldsymbol{v^{\mbox{\tiny \bf +}}}_i} $ must produce equal and opposite components of each ${\boldsymbol{v^{\mbox{\tiny \bf +}}}_k} $ to counterbalance the ${\boldsymbol{v^{\mbox{\tiny \bf +}}}_k} $ components  created by the permittivity distribution.  Physically this means that inhomogeneous or anisotropic, lossy materials induce coupling between eigenmodes.
To demonstrate that this makes at least one pair of eigenmodes nonorthogonal, we make the assumption that all eigenmodes are orthogonal.  Then, substituting Eq.~\ref{eq:inhomogeneous}  into Eq.~\ref{eq:eigenmode equation}, requires that there is a nonzero projection between ${\boldsymbol{v^{\mbox{\tiny \bf +}}}_k} $ and the scattered field of ${\boldsymbol{v^{\mbox{\tiny \bf +}}}_i} $
{\begin{align}
 \int_{V_s} {{{\boldsymbol{v^{\mbox{\tiny \bf +}}}_k}}^*(\boldsymbol{x})} \cdot   \left(  \int _{V_s}  \boldsymbol{\bar{\mathcal{G}}}(\boldsymbol{x},\boldsymbol{x}'){\boldsymbol{v^{\mbox{\tiny \bf +}}}_i}(\boldsymbol{x}') \;{\mathrm{d}^3x'}\right)\mathrm{d}^3x \neq 0 \:.\label{eq:projection}
\end{align}}
We can then consider each ${\boldsymbol{v^{\mbox{\tiny \bf +}}}_i} $ as a linear combination of the eigenmodes $\{ \boldsymbol{g}\} $ (with eigenvalues $\{ \gamma \} $) of the Green's function integral
 {\begin{align}
&{\boldsymbol{v^{\mbox{\tiny \bf +}}}_i}(\boldsymbol{x})= \sum \limits _j f_{ij}\boldsymbol{g_j}(\boldsymbol{x}) \nonumber\\
 & \quad \Rightarrow  \int _{V_s}  \boldsymbol{\bar{\mathcal{G}}}(\boldsymbol{x},\boldsymbol{x}'){\boldsymbol{v^{\mbox{\tiny \bf +}}}_i}(\boldsymbol{x}') \;{\mathrm{d}^3x'} = \sum \limits _j \gamma_j f_{ij}\boldsymbol{g}_j (\boldsymbol{x})\:.
\end{align}}
When neglecting retardation, $  \boldsymbol{\bar{\mathcal{G}}}$ becomes real symmetric and, hence, the eigenmodes $\{ \boldsymbol{g}\}  $ are orthogonal and each coefficient $f_{ij}$ is the complex projection of $\boldsymbol{g}_{j}$ onto ${\boldsymbol{v^{\mbox{\tiny \bf +}}}_i}$.
The orthogonality of $\{ \boldsymbol{g}\} $ in  Eq.~\ref{eq:projection} then requires that there is at least one eigenmode $\boldsymbol{g}_{l}$ that exists in the linear combinations for both ${\boldsymbol{v^{\mbox{\tiny \bf +}}}_i} $ and ${\boldsymbol{v^{\mbox{\tiny \bf +}}}_k} $.
Thus, we can explicitly write the complex inner product between ${\boldsymbol{v^{\mbox{\tiny \bf +}}}_i} $ and ${\boldsymbol{v^{\mbox{\tiny \bf +}}}_k} $  using their respective decompositions into the eigenmodes $\{ \boldsymbol{g}\} $
\begin{align}
\int _{V_s} {\boldsymbol{v^{\mbox{\tiny \bf +}}}_i}^* \cdot \boldsymbol{v}_k\; \mathrm{d}V = \sum \limits_l f_{il}^* f_{kl} \quad (\mathrm{where}\;  f_{il}, f_{kl} \neq 0)\:.
\end{align}
As such, the projection between eigenmodes will generally be nonzero.  For instance, nonorthogonality would be guaranteed if there was only a single shared eigenmode ($\boldsymbol{g}_l$) in both ${\boldsymbol{v^{\mbox{\tiny \bf +}}}_i} $ and ${\boldsymbol{v^{\mbox{\tiny \bf +}}}_k}$.  So, our original assumption, that all eigenmodes in $\{ {\boldsymbol{v^{\mbox{\tiny \bf +}}}_i}\}$ are orthogonal, is broken and we have therefore shown that at least one pair of nondegenerate eigenmodes will be nonorthogonal if the geometry consists of inhomogeneous and/or anisotropic materials.
This also builds on our earlier result (Eq.~\ref{eq:isotropic result}) that circular dichroism in geometries made of isotropic materials can only occur when there are nonorthogonal eigenmodes, because the exception to that condition was the presence of anisotropic materials, which we have just shown leads to nonorthogonal eigenmodes independently.
Therefore we know that nonorthogonal eigenmodes are always necessary for nonreciprocal circular dichroism in absorption.  
Furthermore, we have derived that such nonorthogonal eigenmodes are a result of either retardation of coupling between currents in the structure or from anisotropic or inhomogeneous materials.

It is finally important to acknowledge that small differences in absorption and scattering have recently been observed numerically for a chiral oligomer with planar reflection symmetry~\cite{Ogier2014}.
Additionally, recent experimental observations have also reported a difference in heat generation from a planar chiral structure under excitation by LCP and RCP light~\cite{Shvets2015}.    
These investigations support our derivations on the circular dichroism presented in this pape, but the observed difference in absorption was hypothesized to be a consequence of polarization-dependent near fields, which is not necessarily sufficent to produce a difference in the total absorption as shown in~\cite{HopkinsLiu2013, Rahmani2013}.  
It is nonetheless worth acknowledging that an alternate option is to measure circular dichroism in scattering and absorption from the far-field.  
As demonstrated in Ref.~\cite{Husnik2012}, the scattering and absorption cross sections can be experimentally measured from far-field light by using spatial modulation and interferometry to observe scattering from the interference of scattered field with the incident field.  
It has additionally been demonstrated in Ref.~\cite{Gennaro2014} that measuring the far-field extinction {\it phase}, in addition to amplitude, allows for measurement of the absorption cross section.  
Yet, we should also acknowlege that simpler measurements would be able to observe signs of circular dichroism from the scattering cross section.  
Measurements of transmission or reflection using a high numerical apeture lens have the capacity to capture a large portion of the scattering cross section and will therefore depend on LCP or RCP light.  
Indeed, there are a number of options to experimentally observe the circular dichroism presented here.

\section{Conclusions}

We have presented a rigorous analytical study of circular dichroism in nanostructures with rotational symmetry.
{ It was shown that, because of reciprocity-enforced degeneracies, chiral scattering behavior cannot involve the excitation of modes that are inaccessible (dark) depending on polarization's handedness. }
This observation led to the distinction of two forms of circular dichroism: the traditional form that originates from spatially-distinct excitations and is regularly observed in extinction, and the second form originating from spatially-identical excitations that rotate in opposite directions temporally (such as from reciprocal plane waves) and can be observed in absorption and scattering.
To explain the peculiarities of the second form, we have shown that it can occur only if the scatterer has nonorthogonal eigenmodes. 
Necessary criteria for a scattering structure to have nonorthogonal eigenmodes { was then shown }to be retardation of coupling between currents and/or the use of multiple materials. 
Notably, these will also be necessary criteria for Fano resonances and other such modal interference effects that rely on nonorthogonal eigenmodes. 
A consequence of this analysis is that circular dichroism in absorption would be amplified at locations of significant modal interference, such as Fano resonances.
To demonstrate a manifestation of this circular dichroism effect, we have proposed a planar chiral nanoparticle heptamer that exhibits a Fano resonance and observed that significant absorption circular dichroism occurs in the vicinity of the Fano resonance.
Our conclusions subsequently suggest that there is a key relationship between the modal interference and circularly-dichroic scattering in both linear and nonlinear responses of planar chiral systems.

\section{Acknowledgements}
This work was supported by the Australian Research Council.    ANP acknowledges support of the Russian Foundation for Basic Research and the Deutsche Forschungsgemeinschaft in the frame of International Collaborative Research Center TRR 160.  BH acknowledges the productive discussions with D. A. Powell.

\bibliographystyle{apsrev4-1}
\bibliography{bibliography}

 \end{document}